\documentclass[prd,twocolumn,a4paper,superscriptaddress,floatfix]{revtex4}
\usepackage{graphicx}
\usepackage{bbm}
\usepackage[all]{xy}
\usepackage{amsmath}
\usepackage{amssymb}
\usepackage{epstopdf}

\def\be{\begin{equation}}
\def\ee{\end{equation}}
\newcommand{\bq}{\begin{eqnarray}}
\newcommand{\eq}{\end{eqnarray}}
\newcommand{\bes}{\begin{subequations}}
\newcommand{\ees}{\end{subequations}}

\def\ben{\begin{eqnarray}}
\def\een{\end{eqnarray}}
\def\ba{\begin{array}}
\def\ea{\end{array}}

\begin{document}
\newcommand{\half}{{\textstyle\frac{1}{2}}}
\allowdisplaybreaks[3]
\def\a{\alpha}
\def\b{\beta}
\def\g{\gamma}\def\G{\Gamma}
\def\d{\delta}\def\D{\Delta}
\def\ep{\epsilon}
\def\et{\eta}
\def\z{\zeta}
\def\t{\theta}\def\T{\Theta}
\def\l{\lambda}\def\L{\Lambda}
\def\m{\mu}
\def\f{\phi}\def\F{\Phi}
\def\n{\nu}
\def\p{\psi}\def\P{\Psi}
\def\r{\rho}
\def\s{\sigma}\def\S{\Sigma}
\def\ta{\tau}
\def\x{\chi}
\def\o{\omega}\def\O{\Omega}
\def\k{\kappa}
\def\pa {\partial}
\def\ov{\over}
\def\br{\\}
\def\ud{\underline}

\newcommand\lsim{\mathrel{\rlap{\lower4pt\hbox{\hskip1pt$\sim$}}
    \raise1pt\hbox{$<$}}}
\newcommand\gsim{\mathrel{\rlap{\lower4pt\hbox{\hskip1pt$\sim$}}
    \raise1pt\hbox{$>$}}}
\newcommand\esim{\mathrel{\rlap{\raise2pt\hbox{\hskip0pt$\sim$}}
    \lower1pt\hbox{$-$}}}

\title{Anthropic versus cosmological solutions to the coincidence problem}

\author{A. Barreira}
\email[Electronic address: ]{alex.mr.barreira@hotmail.com}
\affiliation{Centro de F\'{\i}sica do Porto, Rua do Campo Alegre 687, 4169-007 Porto, Portugal}
\affiliation{Departamento de F\'{\i}sica da Faculdade de Ci\^encias
da Universidade do Porto, Rua do Campo Alegre 687, 4169-007 Porto, Portugal}
\author{P.P. Avelino}
\email[Electronic address: ]{ppavelin@fc.up.pt}
\affiliation{Centro de F\'{\i}sica do Porto, Rua do Campo Alegre 687, 4169-007 Porto, Portugal}
\affiliation{Departamento de F\'{\i}sica da Faculdade de Ci\^encias
da Universidade do Porto, Rua do Campo Alegre 687, 4169-007 Porto, Portugal}

\begin{abstract}

In this paper we investigate possible solutions to the coincidence problem in flat phantom dark energy models with a constant dark energy equation of state and quintessence models with a linear scalar field potential. These models  are representative of a broader class of cosmological scenarios in which the universe has a finite lifetime. We show that, in the absence of anthropic constraints, including a prior probability for the models inversely proportional to the total lifetime of the universe excludes models very close to the $\Lambda {\rm CDM}$ model. This relates a cosmological solution to the coincidence problem with a dynamical dark energy component having an equation of state parameter not too close to $-1$ at the present time. We further show, that anthropic constraints, if they are sufficiently stringent, may solve the coincidence problem without the need for dynamical dark energy.

\end{abstract} 
\pacs{98.80.-k, 98.80.Es, 95.35.+d, 12.60.-i}
\maketitle

\section{Introduction}

More than a decade has elapsed since the first observational evidence in favor of cosmic acceleration \cite{Riess:1998cb,Perlmutter:1998np}. Since then, a growing body of independent data has confirmed that our universe is indeed accelerating \cite{Frieman:2008sn,Komatsu:2010fb}. Current observational constraints on the equation of state parameter \cite{Komatsu:2010fb} suggest that a cosmological constant, $\Lambda$, might be playing the role of dark energy. However, the value of $\Lambda$ required to explain the observed cosmic acceleration is off by $\sim120$ orders of magnitude from the standard quantum field theory prediction. Another problem associated with this model is that we seem to be living in a very special epoch where cosmological acceleration has just started. This is known as the coincidence problem.

Dynamical dark energy models alleviate some of the problems associated with the cosmological constant. The so-called tracker quintessence \cite{Zlatev:1998yg,Zlatev:1998tr,Steinhardt:1999nw} and k-essence models \cite{ArmendarizPicon:2000ah,Malquarti:2003hn} consider a scalar field which tracks the behaviour of the dominant component. Also, in interacting quintessence models \cite{PhysRevD.62.043511,Chimento:2003iea,Franca:2003zg,Olivares:2005tb} the dark energy field interacts non-minimally with the matter fields and suitable interaction terms may lead to an attractor scaling solution in which the ratio between dark energy and matter densities remains constant. These models have in common the feature that, except at very early times,  the dynamics of the scalar field responsible for the dark energy is roughly independent of initial conditions. However, they provide no convincent cosmological explanation for the coincidence of our observing time with the onset of cosmological acceleration. Allowing for several accelerating periods has also been considered as a possible solution to the coincidence problem \cite{Dodelson:2001fq,Griest:2002cu,Yang:2005pp}. This way, the present accelerating epoch would be only one of many. However, the major drawback is that there is no evidence that the universe has undergone an accelerating period in the past (other than primordial inflation).

In \cite{Weinberg:1987dv}  an upper bound on $\Lambda$ was obtained, taking into account that larger values would not allow the formation of collapsed structures necessary to accomodate life. This led to the so called multiverse scenario where the cosmological constant is a random variable (one value for each of the universes in the multiverse ensemble). It has been argued that anthropic constraints, in such scenario, select values of $\Lambda$ very close to the observed one \cite{Martel:1997vi,Garriga:1999hu}. In \cite{Egan:2007ht} an estimate was made for the temporal distribution of terrestrial-planet-bound observers which has been used as an argument in favor of an anthropic solution to the coincidence problem (see also \cite{Lineweaver:2003px,Lineweaver:2000da}). However, due to the lack of sufficient understanding of the conditions required for intelligent life to appear, anthropic constraints are usually associated with very large uncertainties.

Cosmological models in which matter and dark energy densities are of the same order of magnitude for a significant fraction of the universe's lifetime avoid the coincidence problem because in such models the present epoch is no longer special. In this paper we consider phantom dark energy models with a constant dark energy equation of state  \cite{Caldwell:2003vq,Scherrer:2004eq} and quintessence models with an effective linear scalar field potential potential \cite{Kallosh:2003bq,Avelino:2004hu,Wang:2004nm,Avelino:2004vy}. These models are representative of a broader class of cosmological scenarios in which the universe has a finite lifetime, thus allowing for a cosmological solution to the coincidence problem if the universe's lifetime is not much larger than the age of the universe today. Anthropic constraints, parametrized by a time cut-off $t_c$ above which observers cannot exist, are also taken into account but we do not attempt  to estimate $t_c$. The main aim of the paper is then to determine  the values of $t_c$ and $w$ for which a solution to the coincidence problem is cosmological, anthropic or both. Throughout the paper we shall assume the metric signature $[+,-,...,-]$ and use units in which $c=\hbar=8\pi G=1$.

\section{Dark energy models}

Consider dark energy models described by a real scalar field $\phi$ minimally coupled to gravity with action
\be\label{model}
S=\int\,d^4x\;{\sqrt{-g}\;\left(-\frac14\,R+{\mathcal L}_m+{\mathcal L(\phi,X)}\right)}\,,
\ee
where ${\mathcal L}_m$ and ${\mathcal L(\phi,X)}$ are, respectively, matter and dark energy Lagrangians, $X=\phi_{,\mu}  \phi^{,\mu}/2$ and a comma represents a partial derivative. If $\phi_{, \mu}$ is timelike then the energy-momentum tensor of the real scalar field can be written in a perfect fluid form
\begin{equation}\label{eq:fluid}
T^{\mu\nu}= (\rho+ p) u^\mu u^\nu - p g^{\mu\nu} \,,
\end{equation}
with
\begin{equation}\label{eq:new_identifications}
u_\mu = \frac{\phi_{, \mu}}{\sqrt{2X}} \,,  \quad \rho = 2 X {\mathcal L}_{,X} - {\mathcal L} \, ,\quad p =  {\mathcal L}(X,\phi)\,.
\end{equation}
In Eq.~(\ref {eq:fluid}), $u^\mu$ is the 4-velocity field describing the motion of the fluid, while $\rho$ and $p$ are its proper energy density and pressure, respectively. The equation of state parameter $w$ is equal to
\begin{equation} 
\label{eq:w}
w \equiv \frac{p}{\rho} = \frac{{\mathcal L}}{2X {\mathcal L}_{,X}  - {\mathcal L}}\,, 
\end{equation} 
and the sound speed squared is given by
\begin{equation}
\label{eq:cs2}
c_s^2 \equiv \frac{p_{,X}}{\rho_{,X}}=\frac{{\mathcal L}_{,X}}{{\mathcal L}_{,X}+2X{\mathcal L}_{,XX}}\,.
\end{equation}
The energy-momentum tensor of the matter field is
\begin{equation}\label{eq:fluidm}
T^{\mu\nu}_{m}= \rho_m v^\mu v^\nu \,,
\end{equation}
where $v^\mu$ is the 4-velocity field of the matter field and $\rho_m$ is its proper energy density. Its proper pressure, $p_m$, 
is equal to zero so that both the equation of state parameter and the sound speed vanish ($w_m=p_m/\rho_m=0$ and $c_{sm}^2=0$).

In a flat homogeneous and isotropic Friedmann-Robertson-Walker (FRW) universe the line element can be written as
\be
\label{metric}
ds^2=dt^2 - a^2(t)\left(dx^2+dy^2 +dz^2\right) \,,
\ee
where $a$ is the scale factor, $t$ is the physical time and $x$, $y$ and $z$ are comoving spatial coordinates. Einstein's equations then imply that
\be
\label{acceleration}
\frac{\ddot a}{a}=-\frac{1}{6}\left(\rho(1+3w)+\rho_m\right) \,,
\ee
where $w=p/\rho$ is the equation of state parameter of the dark energy and a dot represents a derivative with respect to physical time. Energy-momentum conservation for both matter and dark energy components leads to
\bes
\ben
\label{conservm}
{\dot \rho_m} = -3H\rho_m \,,
\\
\label{conserve}
{\dot \rho} = -3H(1+w)\rho\,,
\een
\ees
where $H\equiv {\dot a}/a$ is the Hubble parameter. Eqs. (\ref{conservm}) and (\ref{conserve}) imply that $\rho_m=\rho_{m0} a^{-3}$ while $\rho=\rho_0 a^{-3(1+w)}$ for constant $w$. The subscript `0' refers to the present time which we shall define by the conditions $\Omega_{m0} = 0.27$ and $a_0=1$ throughout the paper. Here $\Omega_m=\rho_m/\rho_c$ and $\rho_c=3H^2$ is the critical density (in a flat universe $\rho_c$ is equal to the total density $\rho_t=\rho_m+\rho$).

\subsection{Model I}

The simplest homogeneous and isotropic phantom dark energy model has a constant equation of state parameter, $w < -1$. In \cite{Caldwell:2003vq} it has been shown what would be the fate of such universe. The energy density of phantom energy increases with time, pushing the universe apart with increasing strength as time goes by. Eventually, the universe would reach the point where any strongly gravitationally bounded objects would be ripped off - such end of the universe has been dubbed a Big Rip. Mathematically, this corresponds to a singularity in the FRW metric scale factor $a(t)$, which becomes infinite in a finite interval of time. The universe's lifetime, is given by \cite{Caldwell:2003vq}
\begin{equation}
t_u \sim  t_0+\frac{2}{3}\frac{1}{\left|1+w\right|H_0\sqrt{1-\Omega_{m0}}}\,.\label{eq:t_U phantom}
\end{equation}

The simplest realization of the above model considers a scalar field Lagrangian with a negative kinetic term given by
\be
\label{phantoml}
{\mathcal L}=-X-V(\phi)\,,
\ee
Dynamically such models are very similar to quintessence models except that, in the phantom case, the scalar field climbs up the potential instead of rolling down. The sound speed squared is equal to unity and 
\begin{equation} 
\label{eq:w1}
w= \frac{-X-V}{-X+V}\,, 
\end{equation}
is smaller than $-1$ as long as $V>X>0$. By requiring that $w<-1$ is a constant, one obtains
\begin{equation}
V=V_{0}a^{-3(w+1)}\label{eq: V(a)}\end{equation}
\begin{equation}
\phi=A+B\ln\left(\frac{a^{3w}}{\left(1+{\sqrt {1+r a^{3w}}}\right)^{2}}\right)\label{eq: phi(a)}\end{equation}
where $B={\sqrt{3|1+w|}}/3w$, $A=B \ln \left(1+{\sqrt{1+r}}\right)^2$  with $r=\Omega_{m0}/\Omega_{e0}$, so that $\phi_0=\phi(a=1)=0$.

Eqs. (\ref{eq: V(a)}) and (\ref{eq: phi(a)}) can be combined to yield the potential as an explicit function of the field
\be
V=V_0\left(\frac{2(1+\sqrt{1+r})\,e^{\frac{\phi}{2B}}}{2(1+\sqrt{1+r})+r(1-e^{\frac{\phi}{B}})}\right)^{-\frac{2(1+w)}{w}}\,.
\label{eq: V(phi)}
\ee
It is possible to show that with this choice of potential the equation of state of dark energy approaches $w$ independently of the initial value of $\dot{\phi}$. The same also holds in the case of the constant $w$ quintessence models investigated in \cite{Sahni:1999gb,2001IJMPD..10..231D,Avelino:2009ze,Avelino:2011ey} (in fact Eqs. (\ref{eq: V(a)}), (\ref{eq: phi(a)}) and (\ref{eq: V(phi)}) also apply in that case). The major drawback of this choice of potential is that it requires a great deal of fine tuning at the transition between the matter and dark energy dominated eras to prevent $w$ to change rapidly there (see \cite{Avelino:2009ze,Avelino:2011ey} for a detailed discussion). For the sake of our analysis, it will be sufficient to take this model as representative of phantom models that lead to a future Big Rip singularity (see \cite{Kujat:2006vj} for a more detailed discussion).

\subsection{Model II}

Consider a quintessence real scalar field  $\phi$ described by the Lagrangian
\be
\label{quintessence}
{\mathcal L}=X-V(\phi)\,,
\ee
where the potential is assumed to be linear with $V(\phi)=V_0+V_1 \phi$ ($V_0$ and $V_1$ are real constants). The assumption of a linear potential is always valid for any realistic model of this type for a certain time interval around the present day.

We shall again assume that $\phi=0$ today so that $V_0$ is the scalar field potential energy at the present time.
The scalar field  
equation of motion is given by
\begin{equation}
\ddot{\phi}+3H\dot{\phi}=-V_1\,,\label{eq:eq motion quinte}
\end{equation}
We shall also assume that the kinetic energy of the field $\phi$ has no memory of initial conditions \cite{Avelino:2004vy}. This appears to be  a reasonable assumption given that, in this model, the memory of initial conditions is rapidly erased.
It has been shown (see \cite{Avelino:2004hu,Avelino:2004vy} for a more detailed description) that this model has a decelerating matter dominated phase followed by a transition to an accelerating dark energy phase. As the field continues to roll down the potential, eventually $V(\phi)$ becomes negative which triggers the rapid collapse of the universe into a Big Crunch. This scenario depends mainly on the linearity of the scalar field potential for $-V_0 \lsim V(\phi) \lsim V_0$ irrespective of the specific form of $V(\phi)$ outside this range. For a given $H_0$, there are only two degrees of freedom, $V_0$ and $V_1$, which correspond to different choices of the parameters $\Omega_{m0}$ and $w$ (we shall reduce the number of degrees of freedom to one by fixing $\Omega_{m0}=0.27$ which essentially fixes $V_0$). We may then use the freedom  to vary $V_1$ (and consequently the value of the equation of state parameter of the dark energy at the present time $w_0$) to numerically compute different possibilities for the total lifetime of the universe $t_u$, using a standard ODE solver. In this model the weighted average value of $w$ in the time interval $[0,t_0]$, 
\begin{equation} 
{\bar w}_0\equiv 
\frac{\int_0^{a_0} da \, (1-\Omega_m) \, w}{\int_0^{a_0} da \, (1-\Omega_m)}\,,
\label{eq8b}
\end{equation}
satisfies $1.6(1+{\bar w}_0) \sim  1+w_0$ \cite{Avelino:2004vy}.

\begin{figure}[t!]
 \includegraphics[width=9cm]{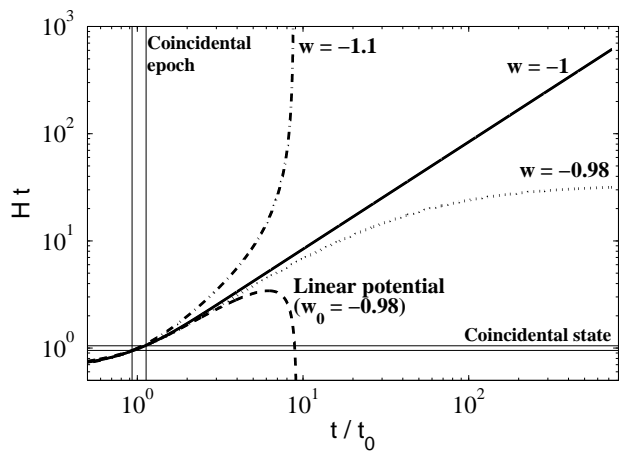}
\caption{The value of $Ht$ as a function of the cosmological time $t$ for four representative cases: $w=-1$ (solid line), $w=-0.98$ (dotted line), model I with $w=-1.1$ (dot-dashed line), and model II with $w_0=-0.98$ (dashed line). The coincidental state, indicated in the figure by the horizontal stripe, was defined by $H t=1\pm0.05$. This defines a coincidental epoch indicated by the vertical stripe.
\label{fig:t vs H^-1-1} }
\end{figure}

\subsection{The age of the universe}

The age of the universe (in units of $H^{-1}$) as a function of $a$ is given by 
\bq
Ht&=&H(a)\intop_{0}^{a}\frac{d {\rm a}}{{\rm a} H({\rm a})}= \nonumber \\
&=&\frac{H(a)}{H_{0}}\int_{0}^{a}\frac{d {\rm a}}{\sqrt{\Omega_{m0}{\rm a}^{-1}+(1-\Omega_{m0}){\rm a}^{-3w-1}}}\,,
\label{ht}
\eq
where $t(a=1)=t_{0}$ and $w$ was assumed to be constant in the last equality of Eq. (\ref{ht}). The value of $Ht$ evolves from being very close to $2/3$ deep into the matter era to $2/(3(w+1))$ at late times if $-1< w \le 0$. If $w\le-1$ then $Ht\to\infty$ as $a\to\infty$ (this happens in a finite timescale if $w<-1$). If dark energy is modelled by a quintessence field with a linear potential, then $Ht$ can no longer be computed using the last equality in Eq. (\ref{ht}) and must be evaluated numerically. In this case, $Ht$ increases from $2/3$ at the start of the dark energy dominated era only to start decreasing  at a later time. The negative  potential energy density of the scalar field then triggers the collapse of the universe with $Ht\rightarrow-\infty$ as $a\rightarrow0$ in a finite time (note that $\dot a < 0$ during the collapsing phase). The fact that the latest observational data is consistent with $H_0 t_0\sim1$ is another way of describing the coincidence problem. This is illustrated in Fig. \ref{fig:t vs H^-1-1} which shows the value of $Ht$ as a function of the cosmological time $t$ for four representative cases: $w=-1$ (solid line), $w=-0.98$ (dotted line), model I with $w=-1.1$ (dot-dashed line) and model II with $w_0=w(a=1)=-0.98$ (dashed line). Here we have taken the conservative estimate $H t=1\pm0.05$ in the definition of coincidental state (horizontal stripe in Fig. \ref{fig:t vs H^-1-1}). This also defines a coincidental epoch, $\Delta t$, indicated by the vertical stripe (note that $\Delta t$ is almost independent of $w_0$ for $w_0 \sim -1$). Fig. \ref{fig:t vs H^-1-1} shows that we seem to be living at a special cosmological time,  $t_0 \sim H_0^{-1}$, very close to the start of the dark energy dominated era. If $w=-1$ (solid line) then $t_u=\infty$ and consequently $\Delta t/t_u =0$. The same applies if $w>-1$ (dotted line). On the other hand, in models I and II the coincidence problem may be strongly alleviated. In these models, the coincidence may be much less dramatic if the universe ends not too far into to the future, either in a Big Rip ($w<-1$, dot-dashed line) or in a Big Crunch (quintessence with linear potential, dashed-line). The coincidental epoch $\Delta t$, corresponding to the coincidental state $H t=1\pm0.05$, may therefore be a significant fraction of the total lifetime of the universe.

\section{Results}

\begin{figure}
\includegraphics[width=9cm]{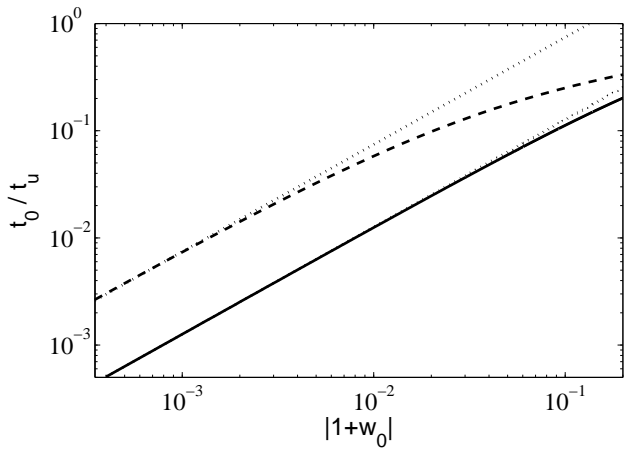}\caption{The value of $t_0/t_u$ as a function of $|1+w_{0}|$ for model I (solid line) and model II (dashed line). The linear approximation (dotted lines) given by Eq. (\ref{eq:tu approx}) holds for $|1+w_{0}|\lesssim10^{-1}$ in the case of model I and $|1+w_{0}|\lesssim10^{-2}$ in the case of model II. \label{fig:linear regimes}}.
\end{figure}

\begin{figure}
\includegraphics[width=9cm]{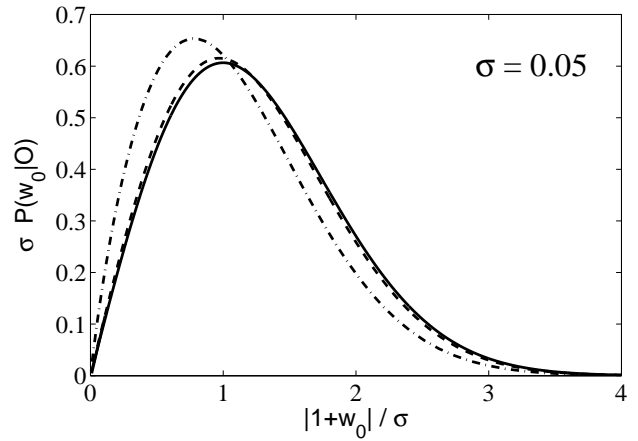}\caption{Posterior probability distributions $P\left(w_{0}|O\right)$ for model I (dashed line) and model II  (dot-dashed line) assuming that  $\sigma=0.05$ and no anthropic constraints. The results are reasonably well described by the analytical approximation given by Eq. (\ref{eq:P_posterior}) (solid line), specially in the case of model I. \label{fig:posterior 0.05}}
\end{figure}

\begin{figure}
\includegraphics[width=9cm]{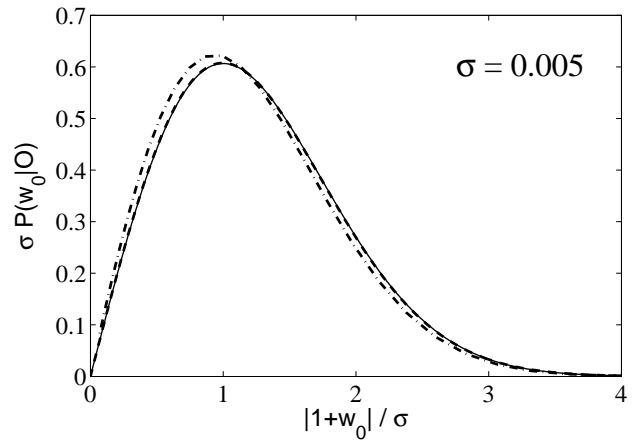}\caption{The same as Fig. \ref{fig:posterior 0.05} but with $\sigma=0.005$. The analytical result  given by Eq. (\ref{eq:P_posterior}) becomes a better approximation for smaller values of $\sigma$.
\label{fig:posterior 0.005}}
\end{figure}

\subsection{Cosmological constraints}

According to the Bayes' theorem 
\begin{equation}
P\left(w_{0}|O\right)=\frac{P\left(O|w_{0}\right)P_{prior}\left(w_{0}\right)}{P(O)}\,,\label{eq:Baye's Theorem}
\end{equation}
where $P_{prior}(w_{0})$ is the prior probability distribution of $w_0$, $P\left(O|w_{0}\right)$ is the probability of the data given the parameter $w_0$ and $P\left(w_{0}|O\right)$ is the posterior probability distribution of $w_0$ (in model I $w=w_0$). The term $P\left(O\right)$ is simply a normalizing constant. In our work, the prior probability in Eq. (\ref{eq:Baye's Theorem}) is assumed to be proportional to the fraction of the total lifetime of the universe corresponding to a coincidental state where $H t= 1 \pm 0.05$. If $t_u$ is not too small then
\begin{equation}
t_u\sim Ct_{0}/|1+w_{0}|\,,\label{eq:tu approx}
\end{equation}
is a good approximation in the case of both models described in the previous section. Here $C>0$ is a model dependent constant of order unity ($C\sim 0.79$ and $C\sim 0.13$ in the case of models I and II, respectively). Hence, 
\begin{equation}
P_{prior}(w_{0}) \propto \frac{\Delta t}{t_u} \propto \frac{1}{t_u} \sim  \frac{1}{t_0} |1+w_0|\,,\label{eq:Prior probability}
\end{equation}
where we used the approximation given by Eq. (\ref{eq:tu approx}). In Fig. \ref{fig:linear regimes}, we plot the value of $t_0/t_u$ as a function of $|1+w_{0}|$ for model I (solid line) and model II (dashed line). The linear approximation (dotted lines) given by Eq. (\ref{eq:tu approx}) holds for $|1+w_{0}|\lesssim10^{-1}$ in the case of model I and $|1+w_{0}|\lesssim10^{-2}$ in the case of model II. For simplicity we shall assume that in model I $P\left(O|w_{0}\right)$ is given by a gaussian distribution with 
\begin{equation}
P(O|w_{0})\propto\exp\left[-\frac{\left(1+w_0\right)^{2}}{2\sigma^{2}}\right]\,,\label{eq:P_observational}
\end{equation}
if $w_{0}\le-1$ and zero otherwise. In model II we shall assume Eq. (\ref{eq:P_observational}) to be valid for $w_0\ge-1$ with $P(O|w_{0})=0$ if $w_0<-1$. Using Eqs. (\ref{eq:Baye's Theorem}), (\ref{eq:Prior probability}) and  (\ref{eq:P_observational}), one obtains the posterior probability distribution
\begin{equation}
P(w_{0}|O)=\frac{|1+w_{0}|}{\sigma^{2}}\exp\left(-\frac{(1+w_{0})^{2}}{2\sigma^{2}}\right)\,,\label{eq:P_posterior}\end{equation}
which satisfies the normalization condition $\int_{-\infty}^{-1}P\left(w_{0}|O\right)d w_0=1$ or $\int_{-1}^{\infty}P\left(w_{0}|O\right)d w_0=1$ in the case of models I or II, respectively. The peak of the probability distribution is at $|1+w_{0}|=\sigma$. Note that the constants of proportionality in Eq. (\ref{eq:Prior probability}), which are model dependent and vary with the definition of coincidental state, are absorbed in the normalization of Eq. (\ref{eq:P_posterior}). In both models, using Eq.  (\ref{eq:P_posterior}), the probability
\begin{equation}
P(|1+w_0|<\Delta)=1-\exp\left(-\frac{\Delta^{2}}{2\sigma^{2}}\right)\sim\frac{\Delta^{2}}{2\sigma^{2}}\,,\label{eq:P_posterior1}
\end{equation}
where the approximation is valid for $\Delta/\sigma\lsim1$.

\begin{figure}
\includegraphics[width=9cm]{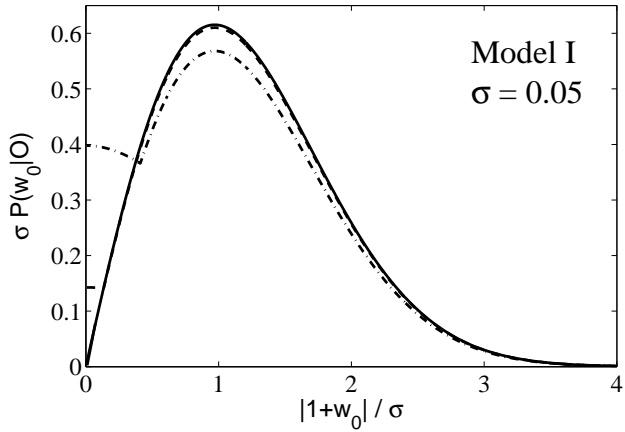}\caption{Posterior probability distributions $P\left(w_{0}|O\right)$ for model I with $t_c=\infty$ (solid line), $t_c/C =50 t_0$ (dot-dashed line) and $t_c/C =150 t_0$ (dashed line), assuming that  $\sigma=0.05$. 
\label{fig:Posteriors with cutoff phantom}}
\end{figure}

Figs. \ref{fig:posterior 0.05} and \ref{fig:posterior 0.005} show the posterior probability distribution as function of $|1+w_{0}|/\sigma$ for models I (dashed lines) and II (dot-dashed lines)  in the absence of anthropic constraints, assuming that $\sigma=0.05$ and $\sigma=0.005$, respectively. Figs. \ref{fig:posterior 0.05} and \ref{fig:posterior 0.005} also show the analytic estimate of the posterior probability obtained using Eq. (\ref{eq:P_posterior}) (solid lines). The quantitative differences registered are expected because for small values of $t_u$ the linear relation in Eq. (\ref{eq:tu approx}), $t_u^{-1} \propto |1+w_0|$, is no longer valid (see Fig. \ref{fig:linear regimes}). The posterior probability distribution $P\left(w_{0}|O\right)$ differs less from Eq. (\ref{eq:P_posterior}) in the case model I than in the case of model II since Eq. (\ref{eq:tu approx}) holds for a wider range of $|1+w_{0}|$ in the phantom model than in the linear quintessence model. Lowering $\sigma$ has the effect of decreasing the statistical importance of the values of $|1+w_{0}|$ for which Eq. (\ref{eq:tu approx}) is not a good approximation and consequently both models have posterior probability distributions  that fit best Eq. (\ref{eq:P_posterior}) with $\sigma=0.005$ than with $\sigma=0.05$, as can be seen by comparing Figs. \ref{fig:posterior 0.05} and \ref{fig:posterior 0.005}. In particular, in Fig.  \ref{fig:posterior 0.005}, the posterior probability distribution of model I (dashed line) almost coincides with the analytical approximation given by Eq. (\ref{eq:P_posterior}) (solid line).

Eqs. (\ref{eq:tu approx}) and (\ref{eq:Prior probability}) show that values of $w_0$ very close to $-1$ lead to very large values of $t_u$ and, consequently, to very small values of the fraction of time where $H t \sim1$. The effect of the prior is therefore to strongly disfavor $w_0=-1$, even when observations that lean towards a cosmological constant are considered. This is clearly seen in Figs. \ref{fig:posterior 0.05} and \ref{fig:posterior 0.005} which show that the peak of the distribution is shifted away from $w_0=-1$. This result links a cosmological solution to the coincidence problem to a significant deviation from the $\Lambda {\rm CDM}$ model. Although models with a constant $w \neq-1$ can be constructed, they do require a large amount of fine tuning \cite{Sahni:1999gb,Avelino:2009ze,Avelino:2011ey} and consequently any deviation from  $w = -1$ should be interpreted as a hint for a dynamical dark energy component.

\begin{figure}
\includegraphics[width=9cm]{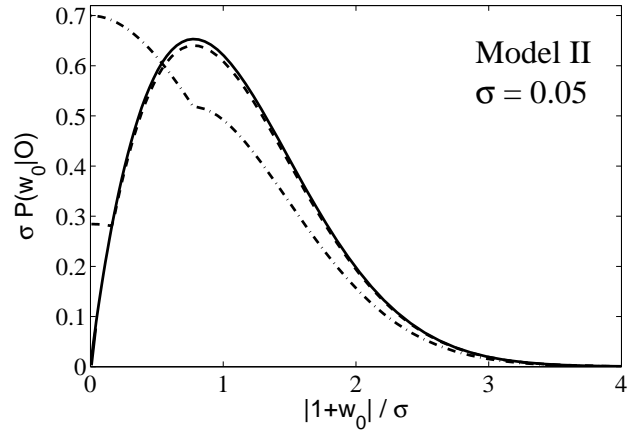}\caption{Posterior probability distributions $P\left(w_{0}|O\right)$ for model II with $t_c=\infty$ (solid line), $t_c/C =50 t_0$ (dot-dashed line) and  $t_c/C =150 t_0$ (dashed line), assuming that  $\sigma=0.05$. 
\label{fig:Posteriors with cutoff quintessence}}
\end{figure}

\subsection{Anthropic constraints}

Anthropic constraints may be added to the analysis by taking into account that there may be time windows for the existence of observers. The corresponding probability distribution function is poorly known (see however \cite{Egan:2007ht}) and consequently in this paper we shall simply assume that there is a cut-off at a time $t_c$ above which observers cannot exist. This way, assuming that the linear approximation given by Eq. (\ref{eq:tu approx}) holds and $C \sim 1$, the prior probability distribution becomes constant for $t_u>t_c$, or equivalently $|1+w_0| \lsim t_0/t_c$. Hence, the posterior probability distribution $P\left(w_{0}|O\right)$ no longer vanishes for $w_0=-1$. On the other hand, for $t_u < t_c$, or equivalently $|1+w_0| \gsim t_0/t_c$, the cut-off has no effect on the shape of the prior probability given by Eq. (\ref{eq:Prior probability}). As a result Eq. (\ref{eq:P_posterior1}) remains valid as long as $t_c \gg  t_0 /\Delta$ but it is no longer a good approximation for $t_c \lsim  t_0 /\Delta$. Unlike the $t_c=\infty$ case the constant $C$ can now affect the results. However, under the transformation $t_c \to t_c/C$, the above discussion remains valid. 

The effect of the anthropic cut-off $t_c$ on the results of models I and II is shown in Figs.  \ref{fig:Posteriors with cutoff phantom} and \ref{fig:Posteriors  with cutoff quintessence}, respectively, which plot the posterior probability distributions $P\left(w_{0}|O\right)$ with $t_c =\infty$ (solid line), $t_c/C =50 t_0$ (dot-dashed line) and  $t_c/C =150 t_0$ (dashed line), for $\sigma=0.05$. One sees that below some value of $|1+w_{0}| \sim C t_0/t_c$, the posteriors aquire the gaussian shape of $P(O|w_{0})$, consequence of the constant prior. If $t_c$ is sufficiently large (dashed lines) then the posterior probability distribution remains essentially unaltered if $|1+w_{0}| \gsim C t_0/t_c$, becoming nearly constant for lower values of $|1+w_{0}|$. On the other hand, for smaller values of $t_c$ (dot-dashed lines) the change of $P(w_0|O)$ for $|1+w_{0}| \lsim C t_0/t_c$ with respect to the $t_c=\infty$ case has a significant impact on the normalization of the probability distribution for $|1+w_{0}|  \gsim C t_0/t_c$.

\begin{figure}
\includegraphics[width=9cm]{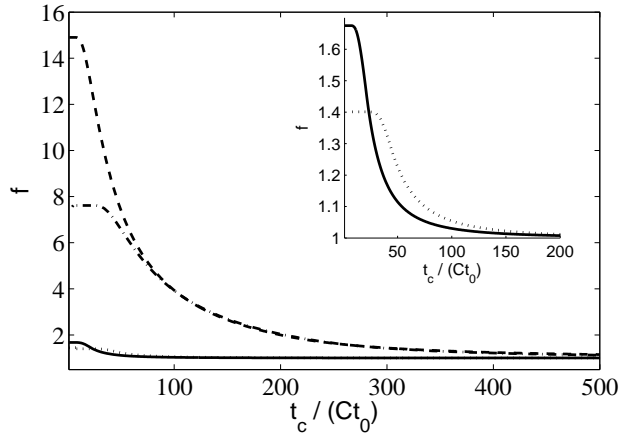}\caption{The value of $f$, defined by Eq. (\ref{eq:f}), as function of $t_c/C$ for models I and II assuming that $\Delta=0.05$ (solid and dotted lines, respectively) and $\Delta=0.005$ (dashed and dot-dashed lines, respectively).\label{fig:impact of t_c}}
\end{figure}

Fig. \ref{fig:impact of t_c} shows the value of 
\be
f=\frac{P(|1+w_{0}|<\Delta)_{t_c}}{P(|1+w_{0}|<\Delta)_{t_c=\infty}}\,,
\label{eq:f}
\ee
as function of $t_c/C$ for models I and II assuming that $\Delta=0.05$ (solid and dotted lines, respectively) and $\Delta=0.005$ (dashed and dot-dashed lines, respectively). For large enough $t_c$ the ratio $f$ approaches unity while for small values of $t_c$ it increases significantly showing that the anthropic constraints may have a large impact on the analysis. For small enough $t_c$ the prior probability distribution becomes independent of $|1+w_{0}|$ which is responsible for the plateaus in Fig. \ref{fig:impact of t_c}. The differences between the results of models I and II for the same $\Delta$, in particular for small $t_c$,  result from the breakdown of the linear approximation given by Eq. (\ref{eq:tu approx}) at larger (smaller) values of $|1+w_0|$ in the case of model I (model II). This is also the reason why, in Figs. \ref{fig:Posteriors with cutoff phantom} and \ref{fig:Posteriors  with cutoff quintessence} for the same value of $t_c/C$, the change of $P(w_0|O)$, with respect to the $t_c=\infty$ case, is more noticeable in the case of model II than in the case of model I.

\section{Conclusion}

In this paper we investigated cosmological as well as anthropic solutions to the coincidence problem in the context of universes  ending in either a Big Crunch or a Big Rip, considering two representative families of cosmological models. In the absence of anthropic constraints, values of $w_0$ very close to $-1$ are ruled out when a prior probability inversely proportional to the total lifetime of the universe is included in the statistical analysis. When anthropic constraints are taken into account, the analysis no longer predict a null probability density for $w_0=-1$. The probability distribution is a function of the assumed time cut-off  $t_c$ for the existence of observers which is uncertain. Depending on the value of $w_0$ and $t_c$ a solution to the coincidence problem can be either anthropic, cosmological or both.


\bibliography{coincidence}

\begin{thebibliography}{33}
\expandafter\ifx\csname natexlab\endcsname\relax\def\natexlab#1{#1}\fi
\expandafter\ifx\csname bibnamefont\endcsname\relax
  \def\bibnamefont#1{#1}\fi
\expandafter\ifx\csname bibfnamefont\endcsname\relax
  \def\bibfnamefont#1{#1}\fi
\expandafter\ifx\csname citenamefont\endcsname\relax
  \def\citenamefont#1{#1}\fi
\expandafter\ifx\csname url\endcsname\relax
  \def\url#1{\texttt{#1}}\fi
\expandafter\ifx\csname urlprefix\endcsname\relax\def\urlprefix{URL }\fi
\providecommand{\bibinfo}[2]{#2}
\providecommand{\eprint}[2][]{\url{#2}}

\bibitem[{\citenamefont{Riess et~al.}(1998)}]{Riess:1998cb}
\bibinfo{author}{\bibfnamefont{A.~G.} \bibnamefont{Riess}} \bibnamefont{et~al.}
  (\bibinfo{collaboration}{Supernova Search Team}), \bibinfo{journal}{Astron.
  J.} \textbf{\bibinfo{volume}{116}}, \bibinfo{pages}{1009}
  (\bibinfo{year}{1998}), \eprint{astro-ph/9805201}.

\bibitem[{\citenamefont{Perlmutter et~al.}(1999)}]{Perlmutter:1998np}
\bibinfo{author}{\bibfnamefont{S.}~\bibnamefont{Perlmutter}}
  \bibnamefont{et~al.} (\bibinfo{collaboration}{Supernova Cosmology Project}),
  \bibinfo{journal}{Astrophys. J.} \textbf{\bibinfo{volume}{517}},
  \bibinfo{pages}{565} (\bibinfo{year}{1999}), \eprint{astro-ph/9812133}.

\bibitem[{\citenamefont{Frieman et~al.}(2008)\citenamefont{Frieman, Turner, and
  Huterer}}]{Frieman:2008sn}
\bibinfo{author}{\bibfnamefont{J.}~\bibnamefont{Frieman}},
  \bibinfo{author}{\bibfnamefont{M.}~\bibnamefont{Turner}}, \bibnamefont{and}
  \bibinfo{author}{\bibfnamefont{D.}~\bibnamefont{Huterer}},
  \bibinfo{journal}{Ann. Rev. Astron. Astrophys.}
  \textbf{\bibinfo{volume}{46}}, \bibinfo{pages}{385} (\bibinfo{year}{2008}),
  \eprint{0803.0982}.

\bibitem[{\citenamefont{Komatsu et~al.}(2011)}]{Komatsu:2010fb}
\bibinfo{author}{\bibfnamefont{E.}~\bibnamefont{Komatsu}} \bibnamefont{et~al.}
  (\bibinfo{collaboration}{WMAP}), \bibinfo{journal}{Astrophys. J. Supp.}
  \textbf{\bibinfo{volume}{192}}, \bibinfo{pages}{18} (\bibinfo{year}{2011}),
  \eprint{1001.4538}.

\bibitem[{\citenamefont{Zlatev and Steinhardt}(1999)}]{Zlatev:1998yg}
\bibinfo{author}{\bibfnamefont{I.}~\bibnamefont{Zlatev}} \bibnamefont{and}
  \bibinfo{author}{\bibfnamefont{P.~J.} \bibnamefont{Steinhardt}},
  \bibinfo{journal}{Phys. Lett.} \textbf{\bibinfo{volume}{B459}},
  \bibinfo{pages}{570} (\bibinfo{year}{1999}), \eprint{astro-ph/9906481}.

\bibitem[{\citenamefont{Zlatev et~al.}(1999)\citenamefont{Zlatev, Wang, and
  Steinhardt}}]{Zlatev:1998tr}
\bibinfo{author}{\bibfnamefont{I.}~\bibnamefont{Zlatev}},
  \bibinfo{author}{\bibfnamefont{L.-M.} \bibnamefont{Wang}}, \bibnamefont{and}
  \bibinfo{author}{\bibfnamefont{P.~J.} \bibnamefont{Steinhardt}},
  \bibinfo{journal}{Phys. Rev. Lett.} \textbf{\bibinfo{volume}{82}},
  \bibinfo{pages}{896} (\bibinfo{year}{1999}), \eprint{astro-ph/9807002}.

\bibitem[{\citenamefont{Steinhardt et~al.}(1999)\citenamefont{Steinhardt, Wang,
  and Zlatev}}]{Steinhardt:1999nw}
\bibinfo{author}{\bibfnamefont{P.~J.} \bibnamefont{Steinhardt}},
  \bibinfo{author}{\bibfnamefont{L.-M.} \bibnamefont{Wang}}, \bibnamefont{and}
  \bibinfo{author}{\bibfnamefont{I.}~\bibnamefont{Zlatev}},
  \bibinfo{journal}{Phys. Rev.} \textbf{\bibinfo{volume}{D59}},
  \bibinfo{pages}{123504} (\bibinfo{year}{1999}), \eprint{astro-ph/9812313}.

\bibitem[{\citenamefont{Armendariz-Picon
  et~al.}(2001)\citenamefont{Armendariz-Picon, Mukhanov, and
  Steinhardt}}]{ArmendarizPicon:2000ah}
\bibinfo{author}{\bibfnamefont{C.}~\bibnamefont{Armendariz-Picon}},
  \bibinfo{author}{\bibfnamefont{V.~F.} \bibnamefont{Mukhanov}},
  \bibnamefont{and} \bibinfo{author}{\bibfnamefont{P.~J.}
  \bibnamefont{Steinhardt}}, \bibinfo{journal}{Phys. Rev.}
  \textbf{\bibinfo{volume}{D63}}, \bibinfo{pages}{103510}
  (\bibinfo{year}{2001}), \eprint{astro-ph/0006373}.

\bibitem[{\citenamefont{Malquarti et~al.}(2003)\citenamefont{Malquarti,
  Copeland, and Liddle}}]{Malquarti:2003hn}
\bibinfo{author}{\bibfnamefont{M.}~\bibnamefont{Malquarti}},
  \bibinfo{author}{\bibfnamefont{E.~J.} \bibnamefont{Copeland}},
  \bibnamefont{and} \bibinfo{author}{\bibfnamefont{A.~R.}
  \bibnamefont{Liddle}}, \bibinfo{journal}{Phys. Rev.}
  \textbf{\bibinfo{volume}{D68}}, \bibinfo{pages}{023512}
  (\bibinfo{year}{2003}), \eprint{astro-ph/0304277}.

\bibitem[{\citenamefont{Amendola}(2000)}]{PhysRevD.62.043511}
\bibinfo{author}{\bibfnamefont{L.}~\bibnamefont{Amendola}},
  \bibinfo{journal}{Phys. Rev. D} \textbf{\bibinfo{volume}{62}},
  \bibinfo{pages}{043511} (\bibinfo{year}{2000}).

\bibitem[{\citenamefont{Chimento et~al.}(2003)\citenamefont{Chimento, Jakubi,
  Pavon, and Zimdahl}}]{Chimento:2003iea}
\bibinfo{author}{\bibfnamefont{L.~P.} \bibnamefont{Chimento}},
  \bibinfo{author}{\bibfnamefont{A.~S.} \bibnamefont{Jakubi}},
  \bibinfo{author}{\bibfnamefont{D.}~\bibnamefont{Pavon}}, \bibnamefont{and}
  \bibinfo{author}{\bibfnamefont{W.}~\bibnamefont{Zimdahl}},
  \bibinfo{journal}{Phys. Rev.} \textbf{\bibinfo{volume}{D67}},
  \bibinfo{pages}{083513} (\bibinfo{year}{2003}), \eprint{astro-ph/0303145}.

\bibitem[{\citenamefont{Franca and Rosenfeld}(2004)}]{Franca:2003zg}
\bibinfo{author}{\bibfnamefont{U.}~\bibnamefont{Franca}} \bibnamefont{and}
  \bibinfo{author}{\bibfnamefont{R.}~\bibnamefont{Rosenfeld}},
  \bibinfo{journal}{Phys. Rev.} \textbf{\bibinfo{volume}{D69}},
  \bibinfo{pages}{063517} (\bibinfo{year}{2004}), \eprint{astro-ph/0308149}.

\bibitem[{\citenamefont{Olivares et~al.}(2005)\citenamefont{Olivares,
  Atrio-Barandela, and Pavon}}]{Olivares:2005tb}
\bibinfo{author}{\bibfnamefont{G.}~\bibnamefont{Olivares}},
  \bibinfo{author}{\bibfnamefont{F.}~\bibnamefont{Atrio-Barandela}},
  \bibnamefont{and} \bibinfo{author}{\bibfnamefont{D.}~\bibnamefont{Pavon}},
  \bibinfo{journal}{Phys. Rev.} \textbf{\bibinfo{volume}{D71}},
  \bibinfo{pages}{063523} (\bibinfo{year}{2005}), \eprint{astro-ph/0503242}.

\bibitem[{\citenamefont{Dodelson et~al.}(2000)\citenamefont{Dodelson,
  Kaplinghat, and Stewart}}]{Dodelson:2001fq}
\bibinfo{author}{\bibfnamefont{S.}~\bibnamefont{Dodelson}},
  \bibinfo{author}{\bibfnamefont{M.}~\bibnamefont{Kaplinghat}},
  \bibnamefont{and} \bibinfo{author}{\bibfnamefont{E.}~\bibnamefont{Stewart}},
  \bibinfo{journal}{Phys. Rev. Lett.} \textbf{\bibinfo{volume}{85}},
  \bibinfo{pages}{5276} (\bibinfo{year}{2000}), \eprint{astro-ph/0002360}.

\bibitem[{\citenamefont{Griest}(2002)}]{Griest:2002cu}
\bibinfo{author}{\bibfnamefont{K.}~\bibnamefont{Griest}},
  \bibinfo{journal}{Phys. Rev.} \textbf{\bibinfo{volume}{D66}},
  \bibinfo{pages}{123501} (\bibinfo{year}{2002}), \eprint{astro-ph/0202052}.

\bibitem[{\citenamefont{Yang and Wang}(2005)}]{Yang:2005pp}
\bibinfo{author}{\bibfnamefont{G.}~\bibnamefont{Yang}} \bibnamefont{and}
  \bibinfo{author}{\bibfnamefont{A.}~\bibnamefont{Wang}},
  \bibinfo{journal}{Gen. Rel. Grav.} \textbf{\bibinfo{volume}{37}},
  \bibinfo{pages}{2201} (\bibinfo{year}{2005}), \eprint{astro-ph/0510006}.

\bibitem[{\citenamefont{Weinberg}(1987)}]{Weinberg:1987dv}
\bibinfo{author}{\bibfnamefont{S.}~\bibnamefont{Weinberg}},
  \bibinfo{journal}{Phys. Rev. Lett.} \textbf{\bibinfo{volume}{59}},
  \bibinfo{pages}{2607} (\bibinfo{year}{1987}).

\bibitem[{\citenamefont{Martel et~al.}(1998)\citenamefont{Martel, Shapiro, and
  Weinberg}}]{Martel:1997vi}
\bibinfo{author}{\bibfnamefont{H.}~\bibnamefont{Martel}},
  \bibinfo{author}{\bibfnamefont{P.~R.} \bibnamefont{Shapiro}},
  \bibnamefont{and} \bibinfo{author}{\bibfnamefont{S.}~\bibnamefont{Weinberg}},
  \bibinfo{journal}{Astrophys. J.} \textbf{\bibinfo{volume}{492}},
  \bibinfo{pages}{29} (\bibinfo{year}{1998}), \eprint{astro-ph/9701099}.

\bibitem[{\citenamefont{Garriga et~al.}(2000)\citenamefont{Garriga, Livio, and
  Vilenkin}}]{Garriga:1999hu}
\bibinfo{author}{\bibfnamefont{J.}~\bibnamefont{Garriga}},
  \bibinfo{author}{\bibfnamefont{M.}~\bibnamefont{Livio}}, \bibnamefont{and}
  \bibinfo{author}{\bibfnamefont{A.}~\bibnamefont{Vilenkin}},
  \bibinfo{journal}{Phys. Rev.} \textbf{\bibinfo{volume}{D61}},
  \bibinfo{pages}{023503} (\bibinfo{year}{2000}), \eprint{astro-ph/9906210}.

\bibitem[{\citenamefont{Egan and Lineweaver}(2008)}]{Egan:2007ht}
\bibinfo{author}{\bibfnamefont{C.~A.} \bibnamefont{Egan}} \bibnamefont{and}
  \bibinfo{author}{\bibfnamefont{C.~H.} \bibnamefont{Lineweaver}},
  \bibinfo{journal}{Phys. Rev.} \textbf{\bibinfo{volume}{D78}},
  \bibinfo{pages}{083528} (\bibinfo{year}{2008}), \eprint{0712.3099}.

\bibitem[{\citenamefont{Lineweaver and Grether}(2003)}]{Lineweaver:2003px}
\bibinfo{author}{\bibfnamefont{C.~H.} \bibnamefont{Lineweaver}}
  \bibnamefont{and} \bibinfo{author}{\bibfnamefont{D.}~\bibnamefont{Grether}},
  \bibinfo{journal}{Astrophys. J.} \textbf{\bibinfo{volume}{598}},
  \bibinfo{pages}{1350} (\bibinfo{year}{2003}), \eprint{astro-ph/0306524}.

\bibitem[{\citenamefont{Lineweaver}(2000)}]{Lineweaver:2000da}
\bibinfo{author}{\bibfnamefont{C.~H.} \bibnamefont{Lineweaver}}
  (\bibinfo{year}{2000}), \eprint{astro-ph/0012399}.

\bibitem[{\citenamefont{Caldwell et~al.}(2003)\citenamefont{Caldwell,
  Kamionkowski, and Weinberg}}]{Caldwell:2003vq}
\bibinfo{author}{\bibfnamefont{R.~R.} \bibnamefont{Caldwell}},
  \bibinfo{author}{\bibfnamefont{M.}~\bibnamefont{Kamionkowski}},
  \bibnamefont{and} \bibinfo{author}{\bibfnamefont{N.~N.}
  \bibnamefont{Weinberg}}, \bibinfo{journal}{Phys. Rev. Lett.}
  \textbf{\bibinfo{volume}{91}}, \bibinfo{pages}{071301}
  (\bibinfo{year}{2003}), \eprint{astro-ph/0302506}.

\bibitem[{\citenamefont{Scherrer}(2005)}]{Scherrer:2004eq}
\bibinfo{author}{\bibfnamefont{R.~J.} \bibnamefont{Scherrer}},
  \bibinfo{journal}{Phys. Rev.} \textbf{\bibinfo{volume}{D71}},
  \bibinfo{pages}{063519} (\bibinfo{year}{2005}), \eprint{astro-ph/0410508}.

\bibitem[{\citenamefont{Kallosh et~al.}(2003)\citenamefont{Kallosh, Kratochvil,
  Linde, Linder, and Shmakova}}]{Kallosh:2003bq}
\bibinfo{author}{\bibfnamefont{R.}~\bibnamefont{Kallosh}},
  \bibinfo{author}{\bibfnamefont{J.}~\bibnamefont{Kratochvil}},
  \bibinfo{author}{\bibfnamefont{A.~D.} \bibnamefont{Linde}},
  \bibinfo{author}{\bibfnamefont{E.~V.} \bibnamefont{Linder}},
  \bibnamefont{and} \bibinfo{author}{\bibfnamefont{M.}~\bibnamefont{Shmakova}},
  \bibinfo{journal}{JCAP} \textbf{\bibinfo{volume}{0310}}, \bibinfo{pages}{015}
  (\bibinfo{year}{2003}), \eprint{astro-ph/0307185}.

\bibitem[{\citenamefont{Avelino et~al.}(2004)\citenamefont{Avelino, Martins,
  and Oliveira}}]{Avelino:2004hu}
\bibinfo{author}{\bibfnamefont{P.~P.} \bibnamefont{Avelino}},
  \bibinfo{author}{\bibfnamefont{C.~J. A.~P.} \bibnamefont{Martins}},
  \bibnamefont{and} \bibinfo{author}{\bibfnamefont{J.~C. R.~E.}
  \bibnamefont{Oliveira}}, \bibinfo{journal}{Phys. Rev.}
  \textbf{\bibinfo{volume}{D70}}, \bibinfo{pages}{083506}
  (\bibinfo{year}{2004}), \eprint{astro-ph/0402379}.

\bibitem[{\citenamefont{Wang et~al.}(2004)\citenamefont{Wang, Kratochvil,
  Linde, and Shmakova}}]{Wang:2004nm}
\bibinfo{author}{\bibfnamefont{Y.}~\bibnamefont{Wang}},
  \bibinfo{author}{\bibfnamefont{J.~M.} \bibnamefont{Kratochvil}},
  \bibinfo{author}{\bibfnamefont{A.~D.} \bibnamefont{Linde}}, \bibnamefont{and}
  \bibinfo{author}{\bibfnamefont{M.}~\bibnamefont{Shmakova}},
  \bibinfo{journal}{JCAP} \textbf{\bibinfo{volume}{0412}}, \bibinfo{pages}{006}
  (\bibinfo{year}{2004}), \eprint{astro-ph/0409264}.

\bibitem[{\citenamefont{Avelino}(2005)}]{Avelino:2004vy}
\bibinfo{author}{\bibfnamefont{P.~P.} \bibnamefont{Avelino}},
  \bibinfo{journal}{Phys. Lett.} \textbf{\bibinfo{volume}{B611}},
  \bibinfo{pages}{15} (\bibinfo{year}{2005}), \eprint{astro-ph/0411033}.

\bibitem[{\citenamefont{Sahni and Starobinsky}(2000)}]{Sahni:1999gb}
\bibinfo{author}{\bibfnamefont{V.}~\bibnamefont{Sahni}} \bibnamefont{and}
  \bibinfo{author}{\bibfnamefont{A.~A.} \bibnamefont{Starobinsky}},
  \bibinfo{journal}{Int. J. Mod. Phys.} \textbf{\bibinfo{volume}{D9}},
  \bibinfo{pages}{373} (\bibinfo{year}{2000}), \eprint{astro-ph/9904398}.

\bibitem[{\citenamefont{{Di Pietro} and {Demaret}}(2001)}]{2001IJMPD..10..231D}
\bibinfo{author}{\bibfnamefont{E.}~\bibnamefont{{Di Pietro}}} \bibnamefont{and}
  \bibinfo{author}{\bibfnamefont{J.}~\bibnamefont{{Demaret}}},
  \bibinfo{journal}{Int. J. Mod. Phys.} \textbf{\bibinfo{volume}{D10}},
  \bibinfo{pages}{231} (\bibinfo{year}{2001}), \eprint{arXiv:gr-qc/9908071}.

\bibitem[{\citenamefont{Avelino et~al.}(2009)\citenamefont{Avelino, Trindade,
  and Viana}}]{Avelino:2009ze}
\bibinfo{author}{\bibfnamefont{P.~P.} \bibnamefont{Avelino}},
  \bibinfo{author}{\bibfnamefont{A.~M.~M.} \bibnamefont{Trindade}},
  \bibnamefont{and} \bibinfo{author}{\bibfnamefont{P.~T.~P.}
  \bibnamefont{Viana}}, \bibinfo{journal}{Phys. Rev.}
  \textbf{\bibinfo{volume}{D80}}, \bibinfo{pages}{067302}
  (\bibinfo{year}{2009}), \eprint{0906.5366}.

\bibitem[{\citenamefont{Avelino et~al.}(2011)\citenamefont{Avelino, Losano, and
  Rodrigues}}]{Avelino:2011ey}
\bibinfo{author}{\bibfnamefont{P.~P.} \bibnamefont{Avelino}},
  \bibinfo{author}{\bibfnamefont{L.}~\bibnamefont{Losano}}, \bibnamefont{and}
  \bibinfo{author}{\bibfnamefont{J.~J.} \bibnamefont{Rodrigues}}
  (\bibinfo{year}{2011}), \eprint{1103.1384}.

\bibitem[{\citenamefont{Kujat et~al.}(2006)\citenamefont{Kujat, Scherrer, and
  Sen}}]{Kujat:2006vj}
\bibinfo{author}{\bibfnamefont{J.}~\bibnamefont{Kujat}},
  \bibinfo{author}{\bibfnamefont{R.~J.} \bibnamefont{Scherrer}},
  \bibnamefont{and} \bibinfo{author}{\bibfnamefont{A.~A.} \bibnamefont{Sen}},
  \bibinfo{journal}{Phys. Rev.} \textbf{\bibinfo{volume}{D74}},
  \bibinfo{pages}{083501} (\bibinfo{year}{2006}), \eprint{astro-ph/0606735}.

\end{thebibliography}

\end{document}